\documentclass[conference]{IEEEtran}
\IEEEoverridecommandlockouts
\usepackage{mathpazo}
\usepackage{times}
\usepackage[latin1]{inputenc}
\usepackage[T1]{fontenc}
\usepackage{dsfont, mathrsfs}
\usepackage{fancyvrb, relsize}
\usepackage{amsmath,amssymb,amsfonts}
\usepackage{algorithmic}
\usepackage{setspace}
\usepackage{algorithm}
\usepackage{subfigure}                                                         
\usepackage{multicol}
\usepackage{graphicx}
\usepackage{epsfig}
\usepackage{upref}
\usepackage{theorem}
\usepackage{tikz}
\usepackage{stfloats}
\usetikzlibrary{arrows}
\usetikzlibrary{positioning}
\usepackage[version=3]{mhchem}
\usepackage{bm}

\let\OLDthebibliography\thebibliography
\renewcommand\thebibliography[1]{
  \OLDthebibliography{#1}
  \setlength{\parskip}{0pt}
  \setlength{\itemsep}{0pt plus 0.8ex}
}

\IEEEoverridecommandlockouts

\theoremstyle{plain} \theorembodyfont{\normalfont\slshape}

\newcommand{\bfc}{{\boldsymbol c}}

\newcommand{\bfg}{{\boldsymbol g}}

\newcommand{\bfm}{{\boldsymbol m}}

\newcommand{\bfs}{{\boldsymbol s}}

\newcommand{\bfu}{{\boldsymbol u}}
\newcommand{\bfv}{{\boldsymbol v}}

\newcommand{\bfx}{{\boldsymbol x}}
\newcommand{\bfy}{{\boldsymbol y}}
\newcommand{\bfz}{{\boldsymbol z}}

\newtheorem{thm}{Theorem$\!$}
\newenvironment{theorem}
{\begin{thm}\hspace*{-1ex}{\bf.}}{\end{thm}}

\newtheorem{lem}[thm]{Lemma$\!$}

\newtheorem{prop}[thm]{Proposition$\!$}
\newenvironment{proposition}{\begin{prop}\hspace*{-1ex}{\bf.}}{\end{prop}}

\newtheorem{cor}[thm]{Corollary$\!$}
\newenvironment{corollary}{\begin{cor}\hspace*{-1ex}{\bf.}}{\end{cor}}

\newtheorem{defn}[thm]{Definition$\!$}
\newenvironment{definition}{\begin{defn}\hspace*{-1ex}{\bf.}}{\end{defn}}

\newtheorem{xmpl}[thm]{Example$\!$}

\newtheorem{cnstr}[thm]{Construction$\!$}
\newenvironment{construction}{\begin{cnstr}\hspace*{-1ex}{\bf.}}{\end{cnstr}}

\begin{document}

\title{Rewriting Flash Memories by Message Passing\vspace{-0.5cm}}
\author{\IEEEauthorblockN{\textbf{Eyal En Gad}, \textbf{Wentao Huang}, \textbf{Yue Li}\ and \textbf{Jehoshua Bruck}}
\IEEEauthorblockA{\emph{California Institute of Technology, Pasadena, CA 91125}\\
\{eengad,whuang,yli,bruck\}@caltech.edu\ \
}\vspace{-1.0cm}
}
%\thanks{This work was supported  by \textbf{FIXME??}.}}
\maketitle

\begin{abstract} This paper constructs WOM codes that combine rewriting and error
correction for mitigating the reliability and the endurance problems
in flash memory. We consider a rewriting model that is of practical
interest to flash applications where only the second write uses WOM
codes.  Our WOM code construction is based on binary erasure
quantization with LDGM codes, where the rewriting uses message passing
and has potential to share the efficient hardware implementations with
LDPC codes in practice. We show that the coding scheme achieves the
capacity of the rewriting model. Extensive simulations show that the
rewriting performance of our scheme compares favorably with that of
polar WOM code in the rate region where high rewriting success
probability is desired. We further augment our coding schemes with
error correction capability. By drawing a connection to the conjugate
code pairs studied in the context of quantum error correction, we
develop a general framework for constructing error-correction WOM
codes. Under this framework, we give an explicit construction of WOM
codes whose codewords are contained in BCH codes.
\end{abstract}

\section{Introduction}
\label{introduction}
Flash memory has become a leading storage media thanks to its many
excellent features such as random access and high storage
density. However, it also faces significant reliability and endurance
challenges. In flash memory, programming cells with lower charge
levels to higher levels can be done efficiently, while the opposite
requires erasing the whole block containing millions of cells. Block
erasure degrades cell quality, and current flash memory can survive
only a small number of block erasures. To mitigate the reliability and
the endurance issues, this paper studies write-once memory (WOM) codes
that combine erasure-free information rewriting and error correction.

\begin{figure}[t]
\begin{center}
\includegraphics[width=0.6\linewidth, angle=270]{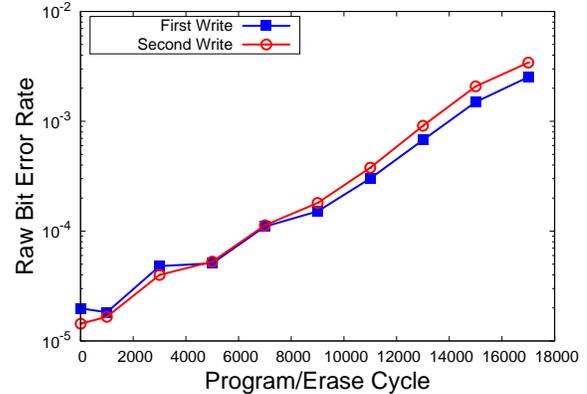}
\end{center}
\vspace{-0.3cm}
\caption{\label{fig:wom-error} The raw BERs when using the proposed
  rewriting scheme.}
\vspace{-0.7cm}
\end{figure}

WOM was first studied by Rivest and Shamir~\cite{RivSha82}. In the
model of WOM, new information is written by only increasing cell
levels. Compared to traditional flash, WOM-coded flash achieves higher
reliability when the same amonut of information is written, or writes
more information using the same number of program/erase (P/E) cycles.
We illustrate these benefits using Fig.~\ref{fig:wom-error}, where we
show the bit error rates (BERs) of the first write and the next
rewrite measured for the scheme of this paper in a 16nm flash chip.
%The figure also shows that the reliability of the second write is
%similar to that of the first write, which validates the feasibility of
%rewriting in flash.
When using the standard setting for error correcting codes (ECCs),
flash memory can survive $14000$ P/E cycles without an ECC decoding
failure. Using a code constructed in this paper that allows user to
write $35\%$ more information, we only need $10370$ P/E cycles to
write the information. Notice that the raw BER at $10370$ P/E cycles
is much lower than that at $14000$ P/E cycles, hence ECC decoding will
have much lower failure rate, which leads to higher reliability. On
the other hand, if we use WOM until ECC fails at $14000$ P/E cycles,
the total amount information that is written requires $18900$ P/E
cycles to write in traditional flash. WOM codes can also be used for \emph{scrubbing} the memory. In this use, the memory is read periodically, to correct errors that were introduced over time. The errors are corrected using an ECC, and the corrected data is written back using a WOM code (see~\cite{LiJiangBru14}). 
Many WOM constructions were
proposed recently. Codes with higher rates were
discovered~\cite{JiaLanSchBru13}\cite{YaaKaySieVarWol12}, and codes
that achieve capacity have also been
found~\cite{BurStr13}.
In this paper, we propose an alternative construction of WOM
codes. Our scheme differs from the WOM codes mentioned above mainly in
two aspects. First, we focus on a specific rewriting model with two
writes, where only the second write uses WOM codes. Such rewriting
scheme has no code rate loss in the first write, and recent
experimental study has demonstrated its effectiveness on improving the
performance of solid state drives~\cite{YadYaaSch15}. Note that, the
model of this rewriting scheme is not only an instance of the general
WOM model~\cite{Hee85}, but also an instance of the model studied by
Gelfand and Pinsker~\cite{GelPin80}. Second, our construction is based
on binary erasure quantization with low-density-generator-matrix
(LDGM) codes. The encoding is performed by iterative quantization
studied by Martinian and Yedidia~\cite{MarYed03}, which is a
message-passing algorithm similar to the decoding of
low-density-parity-check (LDPC) codes. As LDPC codes have been widely
adopted by commercial flash memory controllers, the hardware
architectures of message-passing algorithms have been well understood
and highly optimized in practice. Therefore, our codes are
implementation-friendly for practitioners. Extensive simulations show
that the rewriting performance of our scheme compares favorably with
that of the capacity-achieving polar WOM code~\cite{BurStr13} in the
rate region where a low rewriting failure rate is desired. For
instance, we show that our code allows user to write $40\%$ more
information by rewriting with very high success probability. We note
that the iterative quantization algorithm of~\cite{MarYed03} was used
in~\cite{ChaMarWor06} in a different way for the problem of
information embedding, which share some similarity with our model.

Moreover, our code construction is extended with error correction. The
need for error correction is observed in our experiments. As shown in
Fig.~\ref{fig:wom-error}, the BERs of both writes increase rapidly
with the number of block erasures. 
Constructions of error-correcting WOM codes have been studied in
recent literature. Error-correcting WOM codes have been proposed in~\cite{EngLiKliLanJiaBru14a}\cite{GabSha12}\cite{JiaLiEngLanBru13}\cite{YaaSieVarWol12}\cite{ZemCoh91}. Different
from the existing constructions above, we use conjugate code pairs studied in the context of quantum error
correction~\cite{Ham06}. As an example, we construct LDGM WOM codes whose
codewords also belong to BCH codes. Therefore, our codes allows to use
any decoding algorithm of BCH codes. The latter have been implemented
in most commercial flash memory controllers. We also present two additional approaches to add error correction, and compare their performance.

\section{Rewriting and Erasure Quantization}
\label{sec:rewrite_quant}
%In this section, we define the rewriting model and connects the problem of rewriting to binary erasure quantization. 

%\subsection{Notation}
\subsection{Rewriting Model}
%We consider a rewriting
%model that is of practical interests.
We consider a model that allows two writes on a block of $n$ cells. A cell has a binary state chosen from $\{0,1\}$, with the rewriting constraint that state $1$ can be written to state $0$, but not vice versa. All cells are initially set to be in state 1, and so there is no writing constraint for the first write.
A vector is denoted by a bold symbol, such as $\bfs=(s_1,s_2,\dots,s_n)$.
The state of the $n$ cells after the first write is denoted by the vector $\bfs$.
We focus only on the second write, and we assume that after the first write, the state of the cells is i.i.d., where for each $i$, $\Pr\{s_i=1\} = \beta$. We note that the special case of $\beta=1/2$ is of practical importance, since it approximates the state after a normal page programming in flash memory\footnote{In flash memory, the message to be written can be assumed to be random due to data compression and data randomization used in memory controllers.}. 
%Hence after the first write, the state of the $n$ cells is a random\footnote{In flash memory, the message to be written can be assumed to be random due to data compression and data randomization used in memory controllers.} length-$n$ vector $\bm{s}$ uniformly distributed over $\mathbb{F}_2^n$. T
The second write is concerned with how to store a message $\bm{m} \in \mathbb{F}_2^k$ by changing $\bfs$ to a new state $\bm{x}$ such that 1) the rewriting constraint is satisfied, and 2) $\bm{x}$ represents $\bm{m}$. 
This is achieved by the encoding operation of a rewriting code, defined formally in the following.
\begin{definition}
A \emph{rewriting code} $C_R$ is a collection of disjoint subsets of
$\mathbb{F}_2^n$.
\end{definition}
Each element of $C_R$ corresponds to a different message. Consider $M \in C_R$ that corresponds to a message $\bm{m}$,  then for all $\bm{x} \in M$, we say that $\bm{x}$ is \emph{labeled} by $\bm{m}$. The
decoding function maps the set of labeled vectors into their labels,
which are also the messages. To encode a message $\bfm$ given a state $\bfs$, the encoder needs to find a vector $\bfx$ with
label $\bfm$ that can be written over $\bfs$. If the encoder does not
find such vector $\bfx$, it declares a failure.  The rewriting rate of
$C_R$ is defined by $R_{\mathrm{WOM}}=k/n$. The rewriting capacity, which characterizes the maximum amount of information that 
can be stored per cell in the second write, is known to be $\beta$ bits~\cite{Hee85}.

We are interested in rewriting codes with rates close to the capacity,
together with efficient encoding algorithms with low failure probability. The
main observation in the design of the proposed rewriting scheme of
this paper is that the rewriting problem is related to
the problem of binary erasure quantization (BEQ), introduced in the next subsection. 

\subsection{Binary Erasure Quantization}

The BEQ problem is concerned with the quantization of a binary \emph{source sequence} $\bfs'$, for which some bits are erased. Formally, $\bm{s}' \in \{0,1,*\}^n$, where $*$ represents erasures. $\bfs'$ needs to be quantized (compressed) such that every non-erased symbol of $\bfs'$ will maintain its value in the reconstructed vector. A reconstructed vector with such property is said to have \emph{no distortion} from $\bfs'$. In this paper we use linear BEQ codes, defined as follows:
%For an erasure-source sequence $\bfs'$, let $q(\bfs')$ denote the quantized version of $\bfs'$, according to a code $C_Q$. If $\bfs'$ cannot be quantized with no distortion, we denote $q(\bfs')=F$. Otherwise, $q(\bfs')$ is a codeword of $C_Q$ with no distortion from $\bfs'$.

\begin{definition}
A linear BEQ code $C_Q$ is a subspace of $\mathbb{F}_2^n$. Each $\bfc \in C_Q$ is called a codeword of $C_Q$. The dimension of $C_Q$ is denoted by $r$.
\end{definition}
Each codeword of $C_Q$ is labeled by a different $r$-bits sequence $\bfu$. 
% such that for almost every ``typical'' source sequence $\bfs'$, at least one codeword has no distortion from $\bfs'$. 
Given a BEQ code $C_Q$ and a source sequence $\bfs'$, a quantization algorithm $Q$ is invoked to find a label $\bfu$ whose codeword $\bfc\in C_Q$ has \emph{no distortion} from $\bfs'$. If such label is found, it is denoted by $\bfu=Q(\bfs')$, and is considered as the compressed vector. Otherwise, a quantization failure is declared, and $Q(\bfs')=Failure$. 
%The label $\bfu$ of $Q(\bfs')$ is then declared to be the compressed vector. 
%Since the number of codewords is smaller than the entire space of $n$-bits vector, the labels can be described by less than $n$ bits, and the declared label is actually a compressed vector. 
The reconstruction uses a generator matrix $G_Q$ of $C_Q$ to obtain the codeword $\bfc=\bfu G_Q$. 
%The codeword $Q(\bfs')$ is called the quantized version of $\bfs'$.

\subsection{Reduction from Rewriting to Erasure Quantization}
\label{subset:red}

In this subsection we show that the problem of rewriting can be efficiently reduced to that of BEQ. 
%A BEQ code is linear if it is a subspace of $\mathbb{F}_2^n$. 
Let $C_Q$ be a linear quantization code, and let $H$ be a parity-check matrix of $C_Q$. 
%Given a linear BEQ code $C_Q$, we construct a rewriting code as follows.
\begin{construction}
A rewriting code $C_R$ is constructed as the collection of all cosets of $C_Q$ in $\mathbb{F}_2^n$. 
%The elements of $C_R$ are labeled according to, such that
A decoding function for $C_R$ is defined by a parity check matrix $H$ of $C_Q$, such that a vector $\bfx\in\mathbb{F}_2^n$ is decoded into its syndrome 
\begin{equation}
\label{eq:decode}
\text{DEC}_H(\bfx)=\bfx H^T.
\end{equation}
\end{construction}

 Since the dimension of $C_Q$ is $r$, it has $2^{n-r}$ cosets. Therefore the rate of $C_R$ is $R_{\text{WOM}}=\frac{n-r}{n}$, implying that $k=n-r$. We define some notation before introducing the reduction algorithm. 
Let $(H^{-1})^T$ be a left inverse for $H^T$, meaning that $(H^{-1})^T H^T$ is the $k\times k$ identity matrix. 
%We use $(H^{-1})^T$ as a \emph{coset representative generator}. 
Define a function $BEC: \{0,1\}^n \times \{0,1\}^n \to \{0,1,*\}^n$ as:
\begin{align*}
BEC(\bm{w},\bm{v})_i = \left \{ \begin{array}{cc}
w_i & \text{if }v_i=0\\
* &  \text{if }v_i=1
\end{array}   \right. \ \ \ , \forall i=1,...,n
\end{align*} 
$BEC(\bm{w},\bm{v})$ realizes a binary erasure channel that erases entries in $\bm{w}$ whose corresponding entries in $\bm{v}$ equal 1. We are now ready to introduce the encoding algorithm for the rewriting problem. 

\begin{algorithm}
\caption{$\bfx = ENC(G_{Q}, \bm{m},\bm{s}$): Encoding for Rewriting}
\label{alg:encoding}
%\textbf{Input:} generating matrix $G_{Q}$,  message $\bm{m}$ and  state $\bm{s}$\\
%\textbf{Output:} a vector $\bm{x}$ to rewrite 
\begin{algorithmic}[1]
%\REQUIRE 2) A message vector $\bfm$ of $k$ bits.
%\ENSURE  A vector $\bfx$ labeled by $\bfm$ such that $D(\bfx,\bfs)=0$.
\STATE $\bfz \leftarrow  \bfm (H^{-1})^T $
\STATE $\bfs' \leftarrow BEC(\bfz,\bfs)$
\STATE $\bm{u} \leftarrow Q(\bfs')$
\IF{$\bm{u}$ = FAILURE}
\RETURN FAILURE
\ELSE
\RETURN $\bfx \leftarrow \bm{u}G_Q + \bfz$
\ENDIF
\end{algorithmic}
\end{algorithm} 

%The way to decode and the correctness of Algorithm \ref{alg:encoding} are discussed in Theorem \ref{th:cordec}.
\begin{theorem}\label{th:cordec}
Algorithm \ref{alg:encoding} either declares a failure or returns a vector $\bm{x}$ such that $\bm{x}$ is rewritable over $\bm{s}$ and $\bm{x}H^T = \bm{m}$.
\end{theorem}
\begin{IEEEproof}
Suppose failure is not declared and $\bm{x}$ is returned by Algorithm \ref{alg:encoding}. We first prove that $\bm{x}$ is rewritable over $\bm{s}$. Consider $i$ such that $s_i =0$. Then it follows from the definition of $BEC$ that $s'_i = z_i$. Remember that $Q(\bfs')$ returns a label $\bfu$ such that $\bfc=\bfu G_Q$ has no-distortion from $\bfs'$. Therefore, $c_i=s'_i=z_i$, and $x_i = c_i + z_i = z_i + z_i =0=s'_i$.
So $\bfx$ can be written over $\bfs$. To prove the second statement of the theorem, 
notice that
%by construction,
%\begin{align*}
%\bm{z} H^T = \bfm (H^{-1})^T H^T = \bm{m}.
%\end{align*}
%Because $\bm{c}$ is returned by the quantization algorithm, it is a codeword of $C_Q$ and so $\bm{c} H^T = 0$. 
%Therefore
\begin{align*}
\bm{x}H^T &= (\bm{u}G_Q + \bm{z})H^T = \bm{u}G_QH^T + \bm{m}(H^{-1})^TH^T \\
 &=  \bm{m}(H^{-1})^TH^T = \bm{m}.
\end{align*}
\vspace{-0.2cm}
\end{IEEEproof}

\section{Rewriting with Message Passing}
\label{sec:mp}
In this section we discuss how to choose a quantization code $C_Q$ and quantization algorithm $Q$ to obtain a rewriting scheme of good performance. Our approach is to use the iterative quantization scheme of Martinian and Yedidia~\cite{MarYed03}, where $C_Q$ is an LDGM code, and $Q$ is a message-passing algorithm. 
%From the perspective of implementation, 
This approach is particularly relevant for flash memories, since the hardware architecture of message-passing algorithms is well understood and highly optimized in flash controllers. 

The algorithm $Q$ can be implemented by a sequential or parallel scheduling, as described in~\cite[Section 3.4.2]{MarYed03}. For concreteness, we consider the sequential algorithm denoted by \textsf{ERASURE-QUANTIZE} in~\cite{MarYed03}. Since the performance of \textsf{ERASURE-QUANTIZE} depends on the chosen generator matrix, we abuse notation and denote it by $Q(G_Q,\bfs')$.
%The inputs of \textsf{ERASURE-QUANTIZE} are a generator matrix $G_Q$ and a source $\bfs'$. 
%The output of $\textsf{ERASURE-QUANTIZE}(G_Q,\bfs')$ is a label $\bfu$ of a codeword $\bfc\in C_Q$.
 %Note that $\textsf{ERASURE-QUANTIZE}$ sometimes fail to find a codeword $\bfc$ with no distortion from $\bfs'$.
%Since $Q(G_Q,\bfs')$ needs to be a codeword $\bfc\in C_Q$, we define $Q(G_Q,\bfs')$ to be the product $\bfc=\bfu G_Q$, where $\bfu$ is the output of $\textsf{ERASURE-QUANTIZE}(G_Q,\bfs')$, if it succeed. 
Algorithm $Q(G_Q,\bfs')$ is presented in Appendix~\ref{app:algo}, for completeness.

Finally, we need to describe how to choose a generator matrix $G_Q$ that work well together with Algorithm $Q$. %$G_Q$ should be chosen such that the failure probability of $Q$. 
We show next that a matrix $G_Q$ with good rewriting performance can be chosen to be a \emph{parity-check matrix} that performs well in message-passing decoding of erasure channels. This connection follows from the connection between rewriting and quantization, together with a connection between quantization and erasure decoding, shown in~\cite{MarYed03}. These connections imply that we can use the rich theory and understanding of the design of parity-check matrices in iterative erasure decoding, to construct good generating matrices for rewriting schemes. To make the statement precise, we consider the standard iterative erasure-decoding algorithm denoted by \textsf{ERASURE-DECODE}$(H,\bfy)$ in~\cite{MarYed03}, where $H$ is an LDPC matrix and $\bfy$ is the output of a binary erasure channel. %We denote 
%we use a result of~\cite{MarYed03} that shows a strong connection between the quantization performance of $G_Q$ to the erasure-correction performance 
%Notice that the rate of the LDPC code is $1-{r}/{n}$.

\begin{theorem}
\label{th:ldpc}
For all $\bfm\in\mathbb{F}_2^k$ and $\bfz',\bfs\in\mathbb{F}_2^n$, $\text{ENC}(G_Q,\bfm,\bfs)$ fails if and only if $\textsf{ERASURE-DECODE}(G_Q,\text{BEC}(\bm{z}',\bfs+\bm{1}_n))$ fails, where $\bm{1}_n$ is the all-one vector of length $n$.
%$C_2=C_Q^{\perp}$ fails in the dual BP decoding of a BEC with erasure pattern $\bfs$.
\end{theorem}
The proof of Theorem \ref{th:ldpc} is available in Appendix \ref{app:A}.
The running time of the encoding algorithm ENC is analyzed formally in the following theorem.

\begin{theorem}
\label{th:comp}
The algorithm $\text{ENC}(G_Q,\bfm,\bfs)$ runs in time $\mathcal{O}(nd)$ where $n$ is the length of $\bfs$ and $d$ is the maximum degree of the Tanner graph of $G_Q$.
\end{theorem}
The proof of Theorem \ref{th:comp} is available in Appendix \ref{app:B}.
Theorems~\ref{th:ldpc} and~\ref{th:comp}, together with the analysis and design of irregular LDPC codes that achieve the capacity of the binary erasure channel~\cite{OswSho02}, imply the following capacity-achieving results.

\begin{corollary}
\label{cor:capacity}
There exists a sequence of rewriting codes which can be efficiently encoded by Algorithm~\ref{alg:encoding} and efficiently decoded by Equation \emph{(}\ref{eq:decode}\emph{)} that achieves the capacity of the rewriting model $\beta$.
%The rewriting code $C_R$ achieves the capacity of the rewriting model under message-passing encoding.
\end{corollary}
The proof of Corollary~\ref{cor:capacity} is available in Appendix~\ref{app:capacity}.

\begin{figure}
\begin{center}
\includegraphics[width=0.6\linewidth, angle=270]{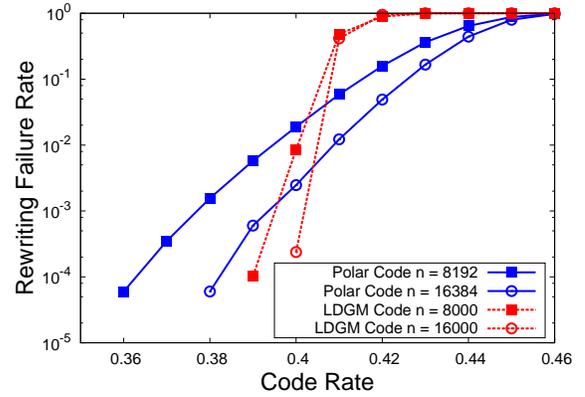}
\end{center}
\vspace{-0.2cm}
\caption{Rewriting failure rates of polar and LDGM WOM codes.}
\label{fig:ldgm-vs-polar}
\vspace{-0.6cm}
\end{figure}

%\subsection{Simulation Results}
The finite-length performance of our rewriting scheme is evaluated
using extensive simulation with the choice of $\beta=0.5$ and $G_Q$ to
be the parity-check matrix of a Mackay code~\cite{Mac99}.  The
rewriting failure rates of our codes with lengths $n = 8000$ and
$16000$ that are relevant to flash applications are compared with
those of the polar WOM codes of lengths $2^{13}$ and
$2^{14}$~\cite{BurStr13}.  Fig.~\ref{fig:ldgm-vs-polar} shows the
rewriting failure rates of both codes at different rewriting rate,
where each point is calculated from $10^5$ experiments. Remember that
the capacity of the model is $0.5$. The results suggest that our
scheme achieves a decent rewriting rate (e.g. 0.39) with low failure
rate (e.g.  $< 10^{-4}$).  Moreover, our codes provide significantly
lower failure rates than polar WOM codes when the rewriting rate is
smaller, because of the good performance in the waterfall region of
message-passing algorithm.

\section{Error-Correcting Rewriting Codes}\label{sec:con}

%This section we propose an approach to combine LDGM rewriting codes with error-correction capability. The proposed approach is based on the construction of conjugate code pair, used in quantum error correction and quantum cryptography. This section is composed of two part. First, in Subsection~\ref{sub:con_frame}, we describe what conjugate code pair are, and to use them as error-correcting rewriting codes. Second, in Subsection~\ref{sub:con_construction}, we present a specific example of conjugate code pair composed of a BCH code for error correction, together with an LDGM code that is used for rewriting.

%\subsection{Conjugate Codes Framework}
%\label{sub:con_frame}

The construction of error-correcting rewriting codes is based on a pair of linear codes $(C_1,C_Q)$, that satisfies the condition $C_1\supseteq C_Q$, meaning that each codeword of $C_Q$ is also a codeword of $C_1$. Define $C_2$ to be the dual of $C_Q$, denoted by $C_2=C_Q^{\perp}$. A pair of linear codes $(C_1,C_2)$, that satisfies $C_1\supseteq C_2^{\perp}$ is called \emph{a conjugate code pair}, and it is useful in quantum error correction and cryptography~\cite{Ham06}. For the flash memory application, we let $C_1$ be an error-correction code, while $C_2^{\perp}=C_Q$ is a BEQ code. The main idea in the construction of error-correcting rewriting codes is to label \emph{only} the codewords of $C_1$, according to their membership in the cosets of $C_Q$. The construction is defined formally as follows:

%We describe in the following how to use a code $C_1$ that contains $C_Q$ as an error-correcting rewriting code.

\begin{construction}
\label{con:conj}
 For $\bfc\in C_1$, let $\bfc+C_Q$ be the coset of $C_Q$ in $C_1$ that contains $\bfc$.
Then the error-correcting rewriting code is constructed to be the collection of cosets of $C_Q$ in $C_1$. 

%by the set $C_1$, where each codeword $\bfx$ of $C_1$ is labeled by $\boldsymbol{\hat{m}}=\bfx H^T$.
\end{construction}
%Consider a linear ECC $C_1$.  limit the rewriting code to contain only codeword of $C_1$. This means that $C_Q$ must be a subcode of $C_1$. To allow a good rewriting performance, $C_Q$ must be the dual of a good erasure-correcting code. We denote the dual of $C_Q$ by $C_2=C_Q^{\perp}$, where $C_2$ is an LDPC code. A pair of codes $(C_1,C_2)$ such that $C_1$ contains the dual of $C_2$ are called \emph{a conjugate code pair}, and are useful also for quantum error-correction and cryptography. We show in Section ?? how to construct conjugate pairs that perform well for the rewriting of flash memory.

%we let the rewriting code be the quotient of a linear ECC by the quantization code $C_Q$. 

%Since an error-correcting rewriting code does not contain the entire space $\mathbb{F}_2^n$, the parity-check matrix of $C_Q$ does not serve well anymore for labeling. Instead, we start by constructing the matrix $(H^{-1})^T$. We consider a pair $(G_1,G_Q)$ of generator matrices of the codes $C_1$ and $C_Q$, respectively, such that each row of $G_Q$ is also a row of $G_1$. Since $C_1$ contains $C_Q$, such matrices pair always exists. We then define $(H^{-1})^T$ to be constructed by the rows of $G_1$ that are \emph{not} rows of $G_Q$.
%This way the coset representative $\bfz$ is a codeword of $C_1$. 
%The decoding algorithm is defined according to the corresponding matrix $H^T$, the right inverse of of $(H^{-1})^T$

%$$\textbf{Error-Correcting Rewriting Code: } C_R \triangleq C_E/C_Q.$$
Next we define the matrices $(H^{-1})^T$ and $H^T$ to be used in encoding and decoding. 
%Consider a conjugate code pair $(C_1,C_Q^{\perp})$. 
Let $G_1$ and $G_Q$ be generator matrices of the codes $C_1$ and $C_Q$, respectively, such that each row of $G_Q$ is also a row of $G_1$. Since $C_1$ contains $C_Q$, such matrix pair always exists. Define $(H^{-1})^T$ to be constructed by the rows of $G_1$ that are \emph{not} rows of $G_Q$. Let $H^T$ be a right inverse of $(H^{-1})^T$.

The encoding is performed according to Algorithm~\ref{alg:encoding}, with the matrix $(H^{-1})^T$ defined above. Note that in Step 1, $\bfz$ is a codeword of $C_1$, since each row of $(H^{-1})^T$ is also a row of $G_1$. In addition, in Step 7, $\bfu G_Q$ is also a codeword of $C_1$ (unless $Q(G_Q,\bfs')$ fails), since $C_Q$ is contained in $C_1$. Therefore, $\bfx=\bfu G_Q+\bfz$ is a codeword of $C_1$. The decoding can begin by the recovery of $\bfx$ from its noisy version, using the decoder of $C_1$. The message $\bfm$ can then be recovered by the product $\bfx H^T$. 

A similar framework was described in~\cite{JacCalSor12}, which proposed a construction of a repetition code contained in a Hamming code, with a Viterbi encoding algorithm. In this paper we make the connection to the quantum coding literature, which allows us to construct stronger codes.

\subsection{Conjugate Codes Construction}
\label{sub:con_construction}

%In this subsection we describe explicit examples of conjugate code pairs. Our intent is to look for constructions with parameters that could be potentially useful in flash memories. 
We look for a conjugate pair $(C_1,C_2)$ such that $C_1$ is a good error-correcting code, while $C_2^{\perp}$ is a good LDGM quantization code. Theorem~\ref{th:ldpc} implies that $C_2$ needs to be an LDPC code with a good performance over a binary erasure channel (under message passing decoding).
%Our focus is on constructions with parameters that could
%Let $C_1$ and $C_2$ be two arbitrary linear codes of the same length. They are said to be conjugate to each other if $C_2^\perp \subset C_1$, or, equivalently, $C_1^\perp \subset C_2$. Pairs of good error correction codes which are conjugate to each other are known to be useful for quantum error correction and quantum cryptography~\cite[p. 2492, last paragraph]{Ste99}
%~\cite{Ham08} [Hamada ITW06, Hamada IT08]. Under our context of rewriting flash memories, pairs of good conjugate error correction codes are also useful, in the following manner. $C_1$ is used in the traditional way to correct errors in $\bf{x}$, i.e., the vector to be rewritten over $\bf{s}$. The dual code of $C_2$, contained in $C_1$, is the BEQ code used in Algorithm 1. The probability of rewriting failure equals the probability of quantization failure of the BEQ code. The latter probability is shown in [Martinian Allerton03] to be determined by the erasure correction capability of $C_2$ with iterative message-passing decoding. Therefore to achieve good rewriting performance, $C_2$ should be chosen to be a LDPC code with good performance over the BEC channel. 
Constructions of conjugate code pairs in which $C_2$ is an LDPC code are studied in~\cite{HagIma07}\cite{IofMez07}\cite{SarKlaRot09}.  Sarvepalli \emph{et al.}~\cite{SarKlaRot09} construct a pair of codes such that $C_1$ is a BCH code and $C_2$ is a Euclidean geometry LDPC code, which is particularly useful for our purpose. This is because BCH codes are used extensively for error correction in flash memories. %, and 2) constructions of low-rate Euclidean geometry LDPC codes are known. 
Below we first briefly review the construction of Euclidean geometry LDPC codes and then discuss the application of the results in~\cite{SarKlaRot09} to our settings. 

\begin{figure}[htbp]
  \begin{center}
      \includegraphics[width=0.3\textwidth, angle=270]{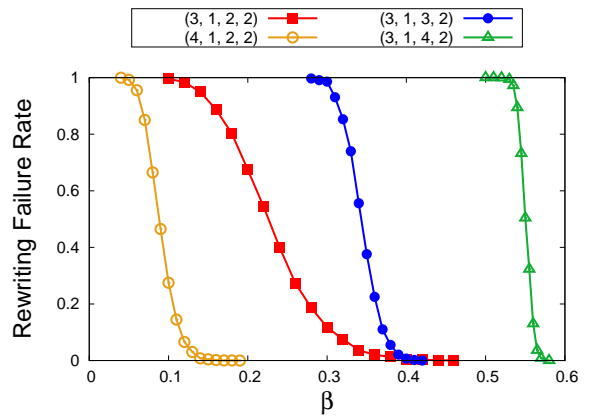}
  \caption{Encoding performance of the codes in Table \ref{womerrcode}.}\label{ach}
  \end{center}
  \vspace{-0.4cm}
\end{figure}

Denote by $\text{EG}(m,p^s)$ the Euclidean finite geometry over $\mathbb{F}_{p^s}$ consisting of $p^{ms}$ points. Note that this geometry is equivalent to the vector space $\mathbb{F}_{p^s}^m$. A $\mu$-dimensional subspace of $\mathbb{F}_{p^s}^m$ or its coset is called a \emph{$\mu$-flat}. Let $J$ be the number of $\mu$-flats that do not contain the origin, and let $\alpha_1, ... \alpha_{p^{sm} - 1}$ be the points of  $\text{EG}(m,p^s)$ excluding the origin. Construct a $J \times p^{sm}-1$  matrix $H_{EG}$ in the way that its $(i,j)$-th entry equals 1 if the $i$-th $\mu$-flat contains $\alpha_j$, and equals 0 otherwise. $H_{EG}$ is the parity check matrix of the (Type-I) Euclidean geometry LDPC code $C_{EG}(m,\mu,s,p)$.  $C_{EG}(m,\mu,s,p)$ is a cyclic code and by analyzing the roots of its generator polynomial, the following result is obtained~\cite{SarKlaRot09}.
\begin{prop}\label{propconj}
$C^\perp_{EG}(m,\mu,s,p)$ is contained in a BCH code of design distance $\delta=p^{\mu s} -1$.
\end{prop}
Hence we may choose $C_2$ to be $C_{EG}(m,\mu,s,p)$ and $C_1$ to be a BCH code with distance equal to or smaller than $\delta$.  Some possible code constructions are shown in Table~\ref{womerrcode}. Their encoding performance, with respect to the probability $\beta$ that a cell in the state is writable, is shown in Fig.~\ref{ach}. Note from Fig.~\ref{ach} that a code with smaller rewriting rate achieves a fixed failure rate at a smaller value of $\beta$. In particular, the codes corresponding to the top three rows of Table~\ref{womerrcode} achieve very small failure rate at $\beta=0.5$, the point of practical interest. These results also show that the slope of the figures becomes sharper when the length of the codes increases, as expected. Out of the three codes that can be rewritten with $\beta=0.5$, $C_{EG}(3,1,3,2)$ poses the best rate and error-correction capability.

\begin{table}
  \caption{Error-correcting Rewriting Codes Constructed from pairs of conjugate BCH and EG-LDPC Codes.}
\vspace{-0.8cm}
\label{womerrcode}
\center
\begin{tabular}{c|c|c|c}
$(m,\mu,s,p)$&$C_1[n,k,\delta]$&$C_2[n,k]$&Rewriting Rate\\
\hline
(4,1,2,2)&[255,247,3]&[255,21]&0.0510\\
(3,1,2,2)&[65,57,3]&[65,13]&0.1111\\
(3,1,3,2)&[511,484,7]&[511,139]&0.2192\\
(3,1,4,2)&[4095,4011,15]&[4095,1377]&0.3158
\end{tabular}
\vspace{-0.5cm}
\end{table}

\section{Alternative Approaches for Error Correction}

In this section we present two alternative approaches to combine rewriting codes with error correction. %We will also give an example of an error-correcting rewriting code constructed using the approach of this section, with concrete performance parameters. 

\subsection{Concatenated Codes}

In this scheme, we concatenate an LDGM rewriting code with a
systematic error-correcting code. The outer code is an LDGM rewriting
code without error-correction capability, as in Section
~\ref{sec:mp}. The systematic ECC is used as the inner code. The
concatenated scheme is used in the second write. The scheme requires
the first write to \emph{reserve} some bits to store the redundancy of
the ECC in the second write.

In the second write, the encoder begins by finding a vector $\bfx$
that can be written over the current state. After $\bfx$ is written,
the systematic ECC calculates the redundancy bits required to protect
$\bfx$ from errors. The redundancy bits are then written into the
reserved cells.  The decoding of the second write begins by recovering
$\bfx$ using the systematic ECC and its redundancy bits. After $\bfx$
is recovered, the decoder of the rewriting code recovers the stored
message from $\bfx$.

We note that reserving bits for the second write have a negative effect on the performance of the system, since it reduces the total amount of information that could be stored in the memory on a given time. Therefore, the next subsection extends the concatenation scheme using a chaining technique, with the aim of reducing the number of bits required to be reserved for the second write.

\subsection{Code Chaining}
The chaining approach is inspired by a similar construction in polar coding~\cite{MonHasUrbSas14}. %The objective of the chaining approach is to reduce the required reserve area without reducing the error-correction performance of the code. 
%Instead, the chaining approach allows to trade-off the rate of the second write. 
%The chaining construction can be seen as an extension of the concatenated construction.
 The idea is to chain several code blocks of short length.  In the following we use a specific example to demonstrate the idea.
 %To demonstrate the idea, we will use specific code parameters. 
 We use a BCH code for error correction, since its performance can be easily calculated. We note, however, that LDPC codes may be used in practice, such that the circuit modules may be shared with the rewriting code, to reduce the required area. The performance of LDPC code in the considered parameters is similar to that of BCH codes.

A typical BCH code used in flash memory has the parameters $[8191,7671,81]$, where the length is $8191$, the dimension is $7671$, and the minimum distance is $81$. If this code is used in a concatenated scheme for the second write, the first write needs to reserve $8191-7671=520$ bits for redundancy.

To reduce the amount of required reserved bits, we consider the chaining of $8$ systematic BCH codes with the parameters $[1023,863,33]$. The encoding is performed sequentially, beginning with the rewriting encoding that finds a vector $\bfx_1$ of $863$ bits. The vector $\bfx_1$ represents a message $\bfm_1$ of $310$ bits, according to an $[863,310]$-LDGM rewriting code. Once $\bfx_1$ is found, the BCH encoder finds $1023-863=160$ redundancy bits to protect $\bfx_1$, as in the concatenated scheme.
%The code is constructed using a systematic $(1023,863,16)$-BCH code, together with an $(863,310)$-LDGM code. The encoding of the second write begins by encoding $310$ information bits into $863$ bits of invalid data. Simulation results show that the encoding failure probability of this code is approximately $1\%$. Let $\bfm_1$ denote the vector of $310$ information bits, and let $\bfx_1$ denote the vector of the $863$ encoded bits. The encoding continues by calculating the $160$ redundancy of $\bfx_1$ in the systematic $(1023,863,16)$-BCH code. 
The encoder then ``chains'' the redundancy bits forward, by encoding them, together with $150$ new information bits, into another block of $863$ bits, using the $[863,310]$-LDGM code. Let $\bfm_2$ denote the vector of $310$ bits encoded into the second block. $\bfm_2$ contains the $160$ redundancy bits of $\bfx_1$, together with the additional $150$ information bits. Note that once $\bfm_2$ is decoded, the redundancy bit of $\bfx_1$ are available, allowing the recovery $\bfx_1$, and then $\bfm_1$. The encoding continues in this fashion $8$ times, to write over a total of $8$ blocks, each containing $863$ cells. The $160$ redundant bits used to protect $\bm{x}_8$ are stored in the reserved cells. The decoding is done in the reverse order, where each decoded vector contains the redundancy bits of the previous block. 

\subsection{Comparison}

We compare the different error-correction approaches, and discuss their trade-offs. The first code in the comparison is a conjugate code pair, described in Section~\ref{sec:con}. We use a conjugation of a $[511,484,7]$-BCH code containing a $[511,372]$-LDGM code, dual to the $(3,1,3,2)$-Euclidean geometry LDPC code in Table~\ref{womerrcode}. The second code in the comparison is a concatenation of an outer $[7671,2915]$-LDGM Mackay rewriting code with an inner $[8191,7671,81]$-BCH code. The third code is a chaining of $8$ blocks of $[863,310]$-LDGM Mackay codes, each concatenated with a $[1023,863,33]$-BCH code. We compare the decoding BER $P_D$, the fraction $\alpha$ of bits required to be reserved, and the rewriting rate $R_{\text{WOM}}$ of the codes. The encoding failure rate of each of the three codes for $\beta=0.5$ is below $10^{-3}$. $P_{D}$ is estimated with a standard flash memory assumption of a raw BER of $1.3\times 10^{-3}$. To achieve a comparable code length, the conjugated code is assumed to be used 16 times in parallel, with a total length of $511\times 16=8176$. 
%The fraction of reserved space in the first write is denoted by $\alpha$. The rewriting code rate is denoted by $R_{\text{WOM}}$. 
The comparison is summarized in Table~\ref{tab:comparison}. 

Flash systems require $P_{D}$ below $10^{-15}$. We see in Table~\ref{tab:comparison} that  conjugated code still do not satisfy the reliability requirement. We also see that concatenated codes that satisfy the reliability requirement need a large fraction of reserved space. The chained code reduces the fraction of reserved space to $2\%$, with a rate penalty in the second write.

%BCH block length$=1023$ RBER = $1.3\times 10^{-3}$ $t=16$, BCH dimension = LDGM block length $= 863$, rewriting code dimension = 310, 8 chaining blocks, probability of LDGM encoding failure $\approx 10^{-3}$ probability of BCH decoding failure $\approx 10^{-16}$. Number of bits to reserve in the first write $= 160$.
%Effective rewriting rate = 
%$$(310 + 7*150)/(7*863+1023)\approx 0.19$$

\begin{table}
  \caption{Error-correcting rewriting codes of length~$\approx 8200$.}
  \vspace{-0.3cm}
\label{tab:comparison}
\center
\begin{tabular}{l | l |c|c}
Code&$P_D$&$\alpha$&$R_{\text{WOM}}$\\
\hline
Conjugated&$10^{-5}$&$0\%$&$0.21$\\
Concatenated&$10^{-16}$&$6.3\%$&$0.35$\\
Chained&$10^{-16}$&$2\%$&$0.19$
\end{tabular}
\vspace{-0.4cm}
\end{table}

\IEEEpeerreviewmaketitle
\allowdisplaybreaks

\bibliographystyle{IEEEtranS}
%\bibliographystyle{unsrt}
%\bibliography{IEEEabrv,local}
\bibliography{local}

\appendices

\section{Iterative Quantization Algorithm}
\label{app:algo}

We denote $G_Q=(\bfg_1,\dots,\bfg_n)$ such that $\bfg_j$ is the $j$-th column of $G_Q$.

\begin{algorithm}
\caption{$\bfu = Q(G_Q,\bfs')$. }
\label{alg:quant}
%\textbf{Input:} A generator matrix $G_Q$ and a source $\bfs'$. \\
%\textbf{Output:} A codeword $\bfc\in C_Q$ with no distortion form $\bfs'$.
\begin{algorithmic}[1]
\STATE $\bfv \leftarrow \bfs'$
%\STATE $\ell \leftarrow 1$
\WHILE {$\exists j$ such that $v_j \ne *$}
	\IF {$\exists i$ such that $\exists! j$ for which $G_{Q}(i,j)  = 1$ and $v_j\ne *$}
		\STATE Push $(i,j)$ into the Stack. %\COMMENT {Reserve $(i,j)$ to later satisfy $x_j = 0$}
	%	\STATE $\ell \leftarrow \ell + 1$
		\STATE $v_j \leftarrow *$.
	\ELSE
		\RETURN FAILURE
	\ENDIF
\ENDWHILE
\STATE $\bfu \leftarrow \bm{0}_{n-k}$ %\COMMENT {The unreserved locations can actually be set arbitrarily.}
%\STATE $\bfz \leftarrow  \bfm C $
\WHILE {Stack is not empty}
	\STATE Pop $(i,j)$ from the Stack.
	\STATE $u_{i} \leftarrow \bfu \cdot \bfg_j + s'_j$ % \COMMENT{Set $u_{r_{\ell,1}}$ to satisfy the rewriting constraint for cell $r_{\ell,2}$. That is, $$x_{r_{\ell,2}}= \bfu \cdot \text{col}_{r_{\ell,2}}G +  \bfm \cdot \text{col}_{r_{\ell,2}}C = 0 $$}
	%\STATE $\ell \leftarrow \ell -1$
\ENDWHILE
\RETURN $\bfu$
\end{algorithmic}
\end{algorithm}

\section{Proof of Theorem \ref{th:ldpc}}\label{app:A}
\begin{IEEEproof}
As in Algorithm~\ref{alg:encoding}, let $\bfz=\bfm (H^{-1})^T$ and $\bfs'=\text{BEC}(\bfz,\bfs)$.
Now according to Algorithm~\ref{alg:encoding}, $\text{ENC}(G_Q,\bfm,\bfs)$ fails if and only if $Q(G_Q,\bfs')$ fails. According to~\cite[Theorem 4]{MarYed03}, $Q(G_Q,\bfs')$ fails if and only if $\text{ERASURE-DECODE}(G_Q,\text{BEC}(\bm{z}',\bfs+\bm{1}_n))$ fails. This completes the proof.
\end{IEEEproof}

\section{Proof of Theorem \ref{th:comp}}\label{app:B}
\begin{IEEEproof}
We first show that Step 1 of Algorithm~\ref{alg:encoding} runs in time $\mathcal{O}(n)$  if $(H^{-1})^T$ is chosen in the following way. For any $C_Q$, its parity check matrix $H$ can be made in to systematic form, i.e., $H = (P\ I)$, by row operations and permutation of columns. Then $(H^{-1})^T$ can be chosen as $(\bm{0}_{k \times n-k}\ I_k)$, and so $\bm{z} = \bm{m}(H^{-1})^T = (\bm{0}_{n-k}\  \bm{m})$.

By \cite[Theorem 5]{MarYed03}, Step 3 of Algorithm~\ref{alg:encoding} runs in time $\mathcal{O}(nd)$. By the definition of $d$, the complexity of Step 7 is also $\mathcal{O}(nd)$.  Therefore $\mathcal{O}(nd)$ dominates the computational cost of the algorithm.
%can be defined as the $k \times n$ matrix $H = [P^T \ I]$. $(H^{-1})^T$ can be chosen as $(\bm{0}_{k \times n-k}, I_k)$.
\end{IEEEproof}

\section{Proof of Corollary~\ref{cor:capacity}}
\label{app:capacity}
\begin{IEEEproof}
Let $\bar{\bfs}=\bfs+\bm{1}_n$. Then it follows from Theorem \ref{th:ldpc} that for all $G_Q$, $\bm{m} \in \mathbb{F}_2^k$, $\bm{z}' \in \mathbb{F}_2^n$, 
\begin{align*}
\Pr \{ &ENC(G_Q,\bm{m},\bm{s}) = Failure  \}  = \\
&  \Pr \{ \textsf{ERASURE-DECODE}(G_Q,\text{BEC}(\bm{z}',\bar{\bfs})) = Failure \},
\end{align*}
where $\bfs$ is distributed i.i.d. with $\Pr\{s_i=\}=\beta$. The right-hand side is the decoding-failure probability of an LDPC code with parity-check matrix $G_Q$ over a binary erasure channel, using message-passing decoding. The erasure probability of the channel is $1-\beta$, because $\Pr\{\bar{s}_i=1\}=1-\Pr\{s_i=1\}$. The capacity of a binary erasure channel with erasure probability $1-\beta$ is $\beta$. This is also the capacity of the rewriting model. In addition, the rate of an LDPC code with parity-check matrix $G_Q$ is equal to the rate of a rewriting code constructed by the cosets of $C_Q$. It is shown in~\cite{OswSho02} how to construct a sequence of irregular LDPC codes that achieves the capacity of the binary erasure channel. Such sequence, used for rewriting codes, achieves the rewriting capacity.
\end{IEEEproof}

\section{Handling Encoding Failures}
The encoding failure event could be dealt with in several ways. A simple solution is to try writing on different invalid pages, if available, or to simply write into a fresh page, as current flash systems do. If the failure rate is small enough, say below $10^{-3}$, the time penalty of rewriting failures would be small. 
For an alternative solution, we state a reformulation of ~\cite[Theorem 3]{MarYed03}.

\begin{proposition}
\label{prop:failure}
For all $\bfm,\bfm'\in\mathbb{F}_2^k$ and $\bfs \in \mathbb{F}_2^n$,
%Let $\bfz',\bfs$ be arbitrary vectors in $\mathbb{F}_2^n$. Then  
 $\text{ENC}(G_Q,\bfm,\bfs)$ fails if and only if $\text{ENC}(G_Q,\bfm',\bfs)$ fails.
%The failure event of $Q(BEC(\bfz,\bfs))$ does not depend on $\bfz$, and is only a function of the state $\bfs$.
\end{proposition}

\begin{IEEEproof}
As in Algorithm~\ref{alg:encoding}, let $\bfz=\bfm (H^{-1})^T$ and $\bfs'=\text{BEC}(\bfz,\bfs)$.
Note that $\text{ENC}(G_Q,\bfm,\bfs)$ fails if and only if $Q(G_Q,\bfs')$ fails. By Algorithm~\ref{alg:quant}, the failure of $Q(G_Q,\bfs')$ is determined only according to the locations of erasures in $\bfs'$, and does not depend on the values of the non-erased entries of $\bfs'$. Since $\bfs'=\text{BEC}(\bfz,\bfs)$, the locations of erasures in $\bfs'$ are only determined by the state $\bfs$. This completes the proof.
\end{IEEEproof}

Proposition~\ref{prop:failure} implies that whether a page is rewritable does not depend on the message to be written. This property suggests that the flash controller can check whether a page is rewritable right after it is being invalidated, without waiting for a message to arrive. An invalid page could be marked as `unrewritable', such that data would be rewritten only into rewritable pages. This policy would guarantee that the rewriting of a new message always succeed. However, this policy also implies that the message passing algorithm would run more than once for the rewriting of a page.

\end{document}